\documentclass[sigconf,nonacm]{acmart}
\acmConference[ICSE 2026]{48th International Conference on Software Engineering}{April 2026}{Rio de Janeiro, Brazil}

\setcopyright{none}
\AtBeginDocument{%
  }

\usepackage{xspace}
\usepackage{cleveref}
\usepackage{listings}
\usepackage{subcaption}
\usepackage{soul}
\usepackage{xcolor}

\usepackage{booktabs}
\usepackage{siunitx}
\usepackage[most]{tcolorbox}
\usepackage{xcolor} 
\usepackage{soul} 

\usepackage[normalem]{ulem}
\usepackage{textcomp}

\usepackage{menukeys}

\newcommand{\MADD}[1]{{\sethlcolor{added}\hl{#1}}} 
\newcommand{\MDELETE}[1]{{\sethlcolor{deleted}\hl{#1}}}
\newcommand{\MCHANGE}[1]{{\sethlcolor{yellow}\hl{#1}}}

\newcommand{\company}{{Google}\xspace}
\newcommand{\gemini}{the LLM\xspace}
\newcommand{\tc}{Transform Code\xspace}
\newcommand{\tool}{AutoPrompter\xspace}

\newcommand{\mysec}[1]{\vspace{0.08cm} \noindent \textbf{#1.}}

\usepackage{ifthen}
\newboolean{showcomments}
\setboolean{showcomments}{true} 
\ifthenelse{\boolean{showcomments}}
  {\newcommand{\nb}[2]{
    \fcolorbox{gray}{yellow}{\bfseries\sffamily\scriptsize#1}
    {\sf\small$\blacktriangleright$\textit{#2}$\blacktriangleleft$}
  }
  
  }
  {\newcommand{\nb}[2]{}
  
  }
  
\newcommand{\eg}{e.g.,\xspace}
\newcommand{\ie}{i.e.,\xspace}

\definecolor{added}{HTML}{AAFFAA}
\definecolor{deleted}{HTML}{FFAAAA}
\definecolor{edited}{HTML}{FFDDB3}

\lstdefinestyle{normal}{
  language=python,
  basicstyle=\ttfamily\scriptsize,
  aboveskip=0pt,
  belowskip=0pt,
  escapeinside={(*}{*)},
}

\lstdefinestyle{a}{
  language=python,
  basicstyle=\ttfamily\scriptsize,
  aboveskip=0pt,
  belowskip=0pt,
  backgroundcolor=\color{added},
  escapeinside={(*}{*)},
}

\newtcbtheorem{Summary}{\bfseries Summary}{enhanced,drop shadow={black!50!white},
  coltitle=black,
  top=0.15in,
  attach boxed title to top left=
  {xshift=1.5em,yshift=-\tcboxedtitleheight/2},
  boxed title style={size=small,colback=white}
}{summary}

\newtcolorbox[auto counter]{prompt}[1][]{title={\bfseries},enhanced,drop shadow={black!50!white},
  coltitle=black,
  top=0.1in,
  attach boxed title to top left=
  {xshift=1.5em,yshift=-\tcboxedtitleheight/2},
  boxed title style={size=small,colback=white},}

\begin{document}

\title{Understanding and supporting how developers prompt for LLM-powered code editing in practice}

\newcommand{\tsc}[1]{\textsuperscript{#1}} 

\makeatletter
\renewcommand{\@authornotemark}{\relax}
\makeatother

\author{Daye Nam\tsc{1*}, Ahmed Omran\tsc{2}, Ambar Murillo\tsc{2}, Saksham Thakur\tsc{2}, Abner Araujo\tsc{2}, Marcel Blistein\tsc{2}, Alexander Fr\"ommgen\tsc{2}, Vincent Hellendoorn\tsc{2}, Satish Chandra\tsc{3*}}

\affiliation{
  \institution{\textsuperscript{1}University of California Irvine \quad \textsuperscript{2}Google \quad \textsuperscript{3}Meta}
  \country{}
}

\authornote{Daye Nam and Satish Chandra conducted this research at Google.}

\renewcommand{\shortauthors}{Daye Nam, et al.}

\begin{abstract}
Large Language Models (LLMs) are rapidly transforming software engineering, with coding assistants embedded in an IDE becoming increasingly prevalent. While research has focused on improving the tools and understanding developer perceptions, a critical gap exists in understanding how developers actually use these tools in their daily workflows, and, crucially, where they struggle. This paper addresses part of this gap through a multi-phased investigation of developer interactions with an LLM-powered code editing feature, \tc, in an IDE widely used at \company. First, we analyze telemetry logs of the feature usage, revealing that frequent re-prompting can be an indicator of developer struggles with using \tc. Second, we conduct a qualitative analysis of unsatisfactory requests, identifying five key categories of information often missing from developer prompts. Finally, based on these findings, we propose and evaluate a tool, AutoPrompter, for automatically improving prompts by inferring missing information from the surrounding code context, leading to a 27\% improvement in edit correctness on our test set.
\end{abstract}

\maketitle

\section{Introduction}

\begin{figure}
\centering
\begin{subfigure}[b]{0.45\textwidth}
  \includegraphics[width=\linewidth]{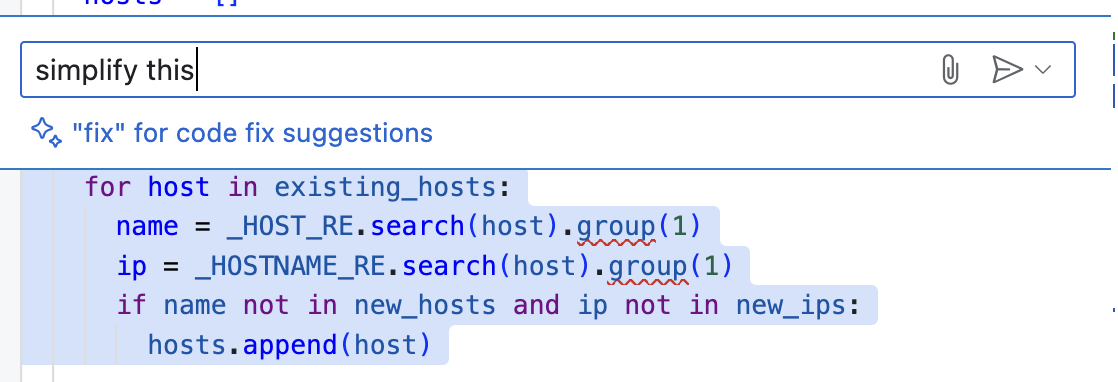}
  \caption{A developer activates \tc using a shortcut or by clicking a button. A \tc prompting window pops up, where the developer can input their prompts.}
  \label{fig:tcbefore} 
\end{subfigure}
\begin{subfigure}[b]{0.45\textwidth}
  \includegraphics[width=\linewidth,trim=0 1cm 0 2cm, clip]{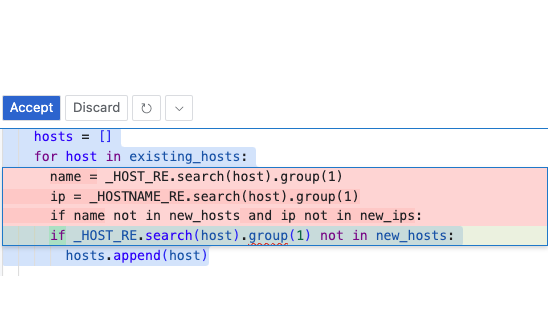}
  \caption{Once the LLM provides a response, the code edit is displayed as diff in the developer's original code. The developer can then either Accept or Discard the suggested code edits (using the first two buttons from the left), or retry the prompt (third button from the left).}
  \label{fig:tcafter}
\end{subfigure}
\vspace{-1em}
\caption{\tc, an LLM-based code editing tool}
\label{fig:tc}
\end{figure}

While the first major win for the use of Large Language Models (LLMs) in coding came from autocomplete~\cite{copilot, ziegler2022productivity}, where the cursor position and the surrounding code were the main inputs to an LLM, the emphasis has shifted to situations in which some amount of developer input -- or \textit{prompt} --- is expected to convey the intent to the LLM.  This holds whether it is for code generation, editing existing code, or as part of a multi-step interaction between the programmer and the LLM.

A notable difference between the autocomplete and prompt-based features is the need for developers to write the prompts.
To receive the desired output from such tools, developers need to clearly elaborate their needs, which can require both time and cognitive effort.
As much previous research on understanding professional developers' perceptions and productivity gains has been conducted with autocomplete as the main focus~\cite{peng2023,paradis2024how,ziegler2022productivity}, we still have unanswered questions on how developers prompt in practice.

\mysec{Transform Code}
We specifically focus on the case of explicitly prompted code editing or transformation, and conduct a case study of \tc (\Cref{fig:tc}).
\tc is a feature in the internal IDE at \company.
Developers can modify and extend code with custom natural language prompts \textit{in context} of a selection of code within the editor, similar to features like GitHub Copilot’s Edit Mode~\cite{copilot} or Cursor's Inline Edit~\cite{cursor}.
It is powered by an LLM trained on many popular programming languages, which is comparable to models like Codex~\cite{chen2021evaluating}.

When a developer selects a certain range of code and activates the feature using a shortcut or by clicking a button, a small \tc window pops up where the developer can enter a short natural language prompt indicating the edit that they would like to see applied to that code.  
Once the LLM provides a response, the suggested code edit is displayed as a diff in the original code, and the developer can either accept or reject it.
For instance, \Cref{fig:tc} shows a request by a developer to \textit{``simplify''} a piece of code (a), and receives a suggested code edit that simplifies the highlighted code (b).  
By providing in-situ support, \tc provides easy contextualization and minimizes context switching.

However, while considerable research has been conducted on prompt engineering for code generation~\cite{chen2025empirical, errica2024did}, mainly for benchmarks such as HumanEval~\cite{chen2021evaluating}, research on the prompting behavior of developers in the enterprise setting remains sparse.
For example, we observed that developers typically use short, simple prompts mostly with a single sentence of around 15 words.
This shows the actual developer prompting patterns are distinct from the code generation benchmark settings, where many prompting engineering approaches utilize a full description of functions with additional information (\eg test cases~\cite{mathews2024test}),

In this paper, we address this gap by investigating how developers interact with LLM-powered code editing tools in practice, how they struggle, and a method for mitigating their struggles through a multi-method, multi-phase investigation.
In Phase I (\Cref{sec:empirical}), we carry out, to our knowledge,  the first \textit{log-based} study of developers' behavior when it comes to prompting for code editing.
With extensive fine-grained logs, we analyze how developers interact with \tc in their daily workflow, with enterprise-level tasks and complexity.
This log-based study minimizes many biases~\cite{dell2012yours} that typically exist with studies asking developers how they feel.
We find that around 7\% of the Transform Code requests are identical re-prompting requests or only minor editing, potentially signaling their struggle in getting a satisfactory response from \tc.

To understand the reasons behind these developers' struggles and propose a mitigation, we build a more focused dataset of unsatisfactory \tc requests  (\Cref{sec:data}).
In addition to the original user prompt and the \tc response, we approximate the ``ground truth'' answer expected by the developer by looking at a future code snapshot or developer-submitted feedback. 
This allows us to ask, \textit{counterfactually}, what would have been a prompt that produced that ground truth answer, or something very close to it.  

Using this data, in Phase II (\Cref{sec:erroranalysis}), we conduct a qualitative study to understand the common issues with the original developer prompts that prevented \tc to generate the ground truth code edit.
We find five common types of information missing from prompts: specifics, operationalization plan, localization/scope, codebase context, and developer intent.

Finally, in Phase III (\Cref{sec:autoprompter}), we test the feasibility of \textit{automatically enhancing the user prompts} to mitigate the user struggles.
We present a prototype tool, AutoPrompter, that considers the common issues of user prompts identified, and automatically augments the user prompts with relevant contextual information.
We find that this tool can improve the effectiveness of developer prompts by around 27\%,  automatically and without further developer input.

\section{Related work}

Although the research in LLM prompting is extensive, we only touch upon the most pertinent pieces of work in AI developer tools and prompting.

\subsection{AI Developer Tools for Professional Developers}

In recent years, driven by significant improvements in large language models (LLMs), many developer tools have been built upon or have incorporated LLMs. 
GitHub Copilot~\cite{copilot} is one of the earliest LLM-powered developer tools, suggesting code in real time, based on context. 
More similar tools have been built, such as Cursor~\cite{cursor}, Windsurf~\cite{windsurf}, and Bolt~\cite{bolt}, supporting code completion, editing, and generation, or even agent-based support.
Significant effort has also been invested in  AI-enhanced software development tools for internal use, such as at Meta~\cite{bader2021} and Google~\cite{AIsoftEng, MLcodecomplete, froemmgen2024}, to support their developers' workflows and proprietary codebases.
Many have conducted studies to understand their potential to enhance developer productivity, investigating both \textit{perceived}~\cite{bird2023, murillo2023} and \textit{actual}~\cite{peng2023, paradis2024how} productivity gains, through various methodologies.
Research has also studied challenges developers face when using AI tools. 
A notable struggle reported multiple times is in effectively communicating requirements to AI tools, which often requires developer cognitive effort~\cite{parameswaran2024revisiting,nam2024understanding,vaithilingam2022,liu2023what,tafreshipour2024prompting}.

Analyzing telemetry logs to understand usage patterns~\cite{murphyhill2018discovering, nam2024understanding} and improve features~\cite{bibaev2022all, froemmgen2024} is a common practice in software engineering. 
However, AI tools have only recently been broadly adopted by developers for real-world usage and are rapidly evolving. 
Consequently, research on how developers incorporate these tools into their actual workflows has been limited. 
More recently, some efforts have been made to measure the overall impact of AI tools at scale, such as by measuring acceptance rates~\cite{ziegler2022productivity} or the number of pull requests~\cite{peng2023}. 
However, these studies primarily focused on quantifying productivity gains rather than understanding \textit{how} developers use the tools. 
One notable exception is \citet{mozannar2024}, which collected and analyzed telemetry data from programmers using Copilot. 
While that paper provides valuable insights into programmer activities using telemetry segments, the data was collected from only 21 participants and the analysis was done after manual labeling. 
In contrast, our analysis examines logs from developers at Google, reporting insights at a large scale.

\subsection{Approaches for Prompt Engineering and Enhancement}

As early research has identified challenges in writing effective prompts \cite{parameswaran2024revisiting,nam2024understanding,vaithilingam2022,liu2023what}, researchers and practitioners have explored ways to assist developers in prompt writing.
Several (semi-) automatic approaches have been proposed, \eg meta-prompting~\cite{yang2024large, fernando2024promptbreeder, zhou2023large} for automatically optimizing prompts in a systematic fashion.
As another example, it was demonstrated in Automatic Prompt Engineer (APE)~\cite{zhou2023large} that LLMs can discover human-level prompts through a search process, outperforming manually written prompts on multiple benchmarks.
However, the main research efforts have focused on optimizing the system instruction prompt that can be used for several distinct tasks within the same category, often requiring many examples to \textit{automatically} quantify the quality gains from prompt improvement.
Another notable example is prompt augmentation tools for image generation, like LMOps~\cite{hao2023optimizing}, which aim at generating diverse images through augmented prompts.
However, their main goal is to generate various outputs, rather than satisfying more constraints by inferring missing specifications from the surrounding context, which is the goal of prompt enhancement for developer tools.

In software engineering domains, researchers have proposed various strategies to mitigate challenges in prompting.
Several attempts have been made to enhance code generation quality through techniques such as adding test cases~\cite{mathews2024test}, planning mechanisms~\cite{jiang2024self}, or leveraging semantic entropy distributions of generated programs~\cite{jia2025automated}. However, these approaches primarily focus on optimizing code generation performance by encouraging LLMs to engage in more sophisticated reasoning, rather than supporting the diverse code editing tasks that developers encounter in practice.
Other researchers have proposed proactively providing assistance to reduce the cognitive burden on developers during prompt engineering~\cite{nam2024understanding, need2025chen}. 
These approaches are useful when users are unfamiliar with a domain or when their intent is easily inferred. However, it is unclear whether those benefits can be generalized in enterprise environments, where developers work on complex, specialized tasks that make intent inference substantially more difficult.
Some studies have explored enhancing initial prompts by requesting clarification questions~\cite{li2022python}, which can be highly effective for correctly inferring user intent. However, this approach imposes cognitive overhead on users, who must read follow-up questions and provide appropriate responses before receiving the desired output.
To the best of our knowledge, ours represents the first study to analyze the prompt strategies employed by professional developers in their daily workflows, while investigating the feasibility of directly enhancing user prompts to minimize cognitive load from multi-turn clarification or debugging processes.

\section{Phase 1: \tc Log Analysis}
\label{sec:empirical}

\tc has been adopted by a significant fraction of developers since its release.
Through iterative improvements to both the underlying LLM and the user interface, \tc has become a core part of developers' workflows. 
Despite its positive reception and appreciation, generating code edits from natural language prompts remains a relatively new concept for developers. 
Thus, some developers have expressed challenges, mirroring the experiences of users of similar tools~\cite{kruse2024can, nam2024understanding}. 

In the ideal case, a developer using \tc would send a request with a prompt for a given task, \tc would provide an appropriate code edit, the developer would review the edit, and proceed to the next task. However, in practice, it is unrealistic to expect an LLM-powered tool to provide the perfect solution on the first attempt. 
In this section, we analyze telemetry logs from \tc and the IDE, focusing on how developers use \tc, and how they react when \tc does not provide the ideal code edit they anticipated.

\subsection{Data}

To understand developer struggles and think of possible remedies, we have to start with data.  Here is the data that we have available based on actual usage of \tc.

\subsubsection{In-IDE Event Logs}
We collected one month of user telemetry logs from the internal \company IDE, from January 27, 2025, to February 23, 2025. 
The telemetry data, which is collected automatically, contains in-IDE events ranging from feature usage (e.g., \tc) to basic actions (e.g., pressing the Enter key), along with timestamps for each event. 

\subsubsection{\tc Logs}
Logs for specific features covering non-IDE aspects, \eg \tc prompts, are collected separately at \company.
\tc-specific telemetry logs capture developer interactions with \tc, including timestamps, prompts, the context in which \tc was triggered, any code highlighted by the developer, and the generated \tc edits.

\subsubsection{Dataset used for analysis}
From this combined dataset, we sampled data of over 10,000 individual developers who used \tc at least 24 times during the month, which is the median usage of the entire population. 
This threshold was set to minimize noise arising from non-regular usage of \tc, \eg where new users might test \tc with simple prompts like \textit{"hi."}

\subsubsection{Limitations}

The IDE was instrumented to log an event whenever developers performed any edit, along with the source of this edit (e.g., copy/pasting, typing, running a code formatter, etc.). 
Due to the number of possibilities for how developers can perform edits in the IDE, a small percentage of edits (<1\%) do not have any source assigned to them. 
Many data points were also excluded during data preprocessing, for example, when joining in-IDE event logs and \tc logs, as the logs are collected at different granularities.
However, given the large size of the data, we believe it still shows the general workflow when \tc is available.

\begin{figure}[t]
    \centering
    \begin{subfigure}[b]{0.45\textwidth}
    \centering
  \includegraphics[width=0.75\linewidth]{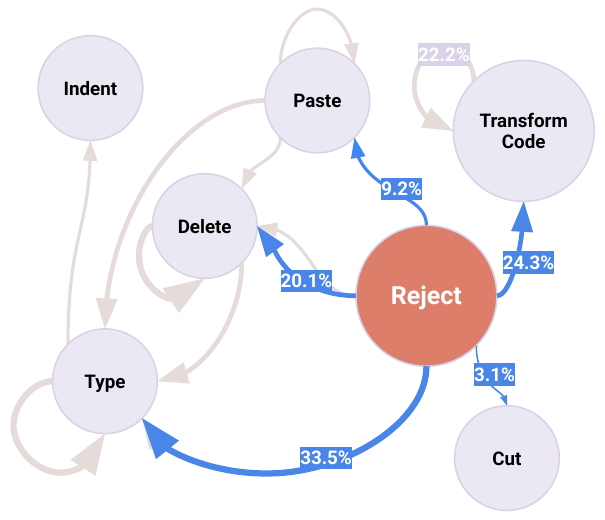}
  \caption{After rejection.}
\end{subfigure}
    \begin{subfigure}[b]{0.45\textwidth}
    \centering
  \includegraphics[width=0.75\linewidth]{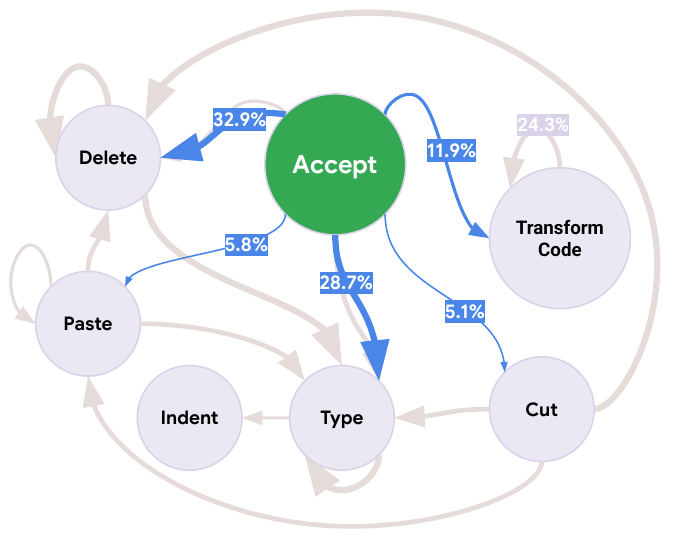}
  \caption{After acceptance.}
\end{subfigure}
    \caption{In-IDE developer action transition graphs. Each node displays an IDE action, and each edge indicates the proportion of transitions between the connected actions. Blue arrows show the first action taken right after a \tc request, and gray arrows indicate the second action following the first. For space and readability reasons, edges representing transition frequencies lower than 3\% (for the first action) and 10\% (for the second action) have been omitted.}
    \label{fig:transition}
\end{figure}

\subsection{Post-\tc Interactions}
\label{subsec:posttc}

To first understand how developers interact with \tc in real-world tasks, we focused on the in-IDE actions developers take after their \tc requests.
We collected the subsequent in-IDE developer actions after requesting \tc and either accepting or rejecting a response.
The challenge here is to accurately associate whether a certain action is a reaction to another, as we cannot reliably determine the developer's intent without asking them.
Thus, as a best approximation, we only included the \tc requests and the subsequent in-IDE actions, when they happened within a short window, 10 minutes, and when the subsequent in-IDE actions were made in the same parts of the code.

\Cref{fig:transition} visualizes the two consecutive in-IDE developer actions after requesting \tc.
We first analyzed the actions made after rejection, \Cref{fig:transition}-(a), to understand how developers cope with unsatisfactory \tc edits.
Then, we also analyzed the actions made after acceptance, \Cref{fig:transition}-(b), to check for any differences in developers' reactions.
This distinction is also important because the \tc edit is applied to the user's code upon acceptance, but discarded upon rejection.
Thus, subsequent edits made by the user in the same parts of the code after acceptance likely indicate refinement or adjustment of the applied \tc edit, whereas actions taken after rejection indicate the user is continuing to work on the original code.

\subsubsection{Post-rejection}
The most common subsequent action after rejecting a \tc edit was Type (33.5\%), including typing a character or inserting a line.
This indicates that developers often start working on their tasks by themselves after checking an unsatisfactory \tc edit.
A particularly interesting next action was a repeated request to \tc (24.3\%), which is the second most common action after rejecting a \tc edit.
This likely indicates that developers re-prompted \tc for the same task.
We found that not an insignificant proportion of developers (22.2\%) made even another \tc request, possibly indicating developer struggle in using \tc, and/or their desire to delegate the task to \tc instead of picking up on the task themselves, like those who started typing.
The following frequent subsequent actions were Delete (20.1\%), Paste (9.2\%), and Cut (3.1\%), showing other mechanisms of editing the original code, possibly adopting code from code or documentation search.

\subsubsection{Post-acceptance}
The most common subsequent action was Delete (32.9\%), and the second most frequent action was Type (28.7\%).
This indicates that the developers frequently make edits \textit{to} the \tc-edited code, after accepting a response, to further refine it.
The next most frequent action was a repeated request to \tc (11.9\%). 
Even after accepting the \tc edit, users requested \tc from the same code region more than 10\% of the time.
This could indicate that developers found unsatisfactory edits \textit{after} accepting the edits, and re-prompted for the same task, or it is still possible that developers moved on to the next task and requested another code edit within the same part (e.g., adding documentation after implementing a function). 
Developers also made edits via other mechanisms, possibly to the \tc-edited code, by pasting code (5.8\%), or cutting the code (5.1\%).

\subsection{Characteristics of Re-prompting}
By analyzing the post-\tc interactions, we found that developers often request \tc again without making any changes to the code (\textit{re-prompting}).
This either indicates that \tc provided a code edit that perfectly addressed the user's request and the user moved on to the next task, or, the suggested code edit was unsatisfactory and the user made a follow-up request.
To understand this signal better, we conducted a more focused analysis on consecutive \tc requests.

\subsubsection{Frequency of Identical Re-prompting}
We first investigated how often developers re-prompt with an identical prompt.
We started with identical re-prompting, as it is very likely that the user is making a follow-up request, instead of moving on to the next task after getting a satisfactory response.

In this analysis, we considered a request as an identical re-prompting, when it uses the identical prompt as the previous request and the request is made within 10 minutes after the previous request.
As one may use the same prompt for different code locations for different tasks -- for example, one can use \textit{``write javadoc''} for different functions, which indicates that the user is satisfied with the \tc edits -- we also excluded cases where the code selected for the request (e.g., highlighted code in \Cref{fig:tc} (a)) was distant (not within 3 lines).

Out of a set of $\sim$100,000 \tc requests, we found that roughly 7\% of requests repeated the previous prompt.
There were even some (rare) cases where a developer re-prompted \tc with the identical prompt up to 13 times. 
This indicates that a significant portion of consecutive \tc requests are indeed follow-up requests for an unsatisfying code edit, rather than developers moving on to the next task after receiving a satisfactory code edit.

\subsubsection{Re-prompting with tweaks}

\begin{figure}
\centering
\begin{subfigure}[b]{0.45\textwidth}
  \includegraphics[width=1\linewidth]{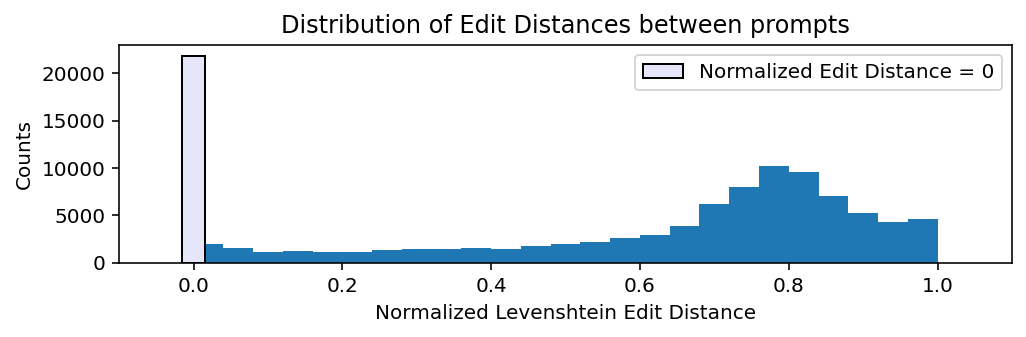}
\end{subfigure}
\begin{subfigure}[b]{0.45\textwidth}
  \includegraphics[width=1\linewidth]{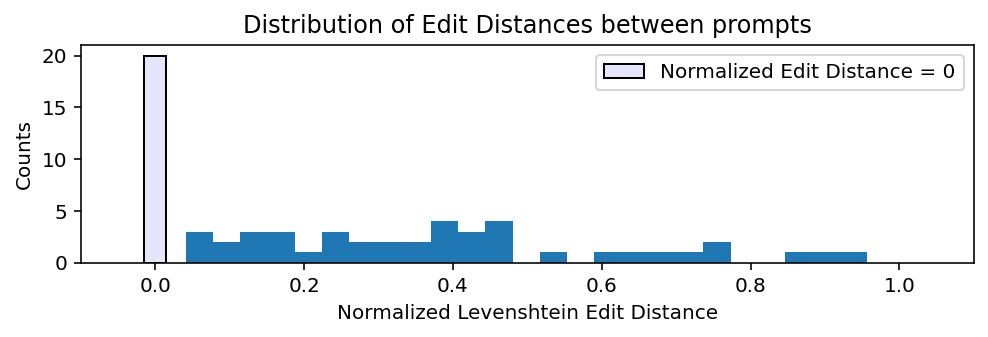}
\end{subfigure}
\begin{subfigure}[b]{0.45\textwidth}
  \includegraphics[width=1\linewidth]{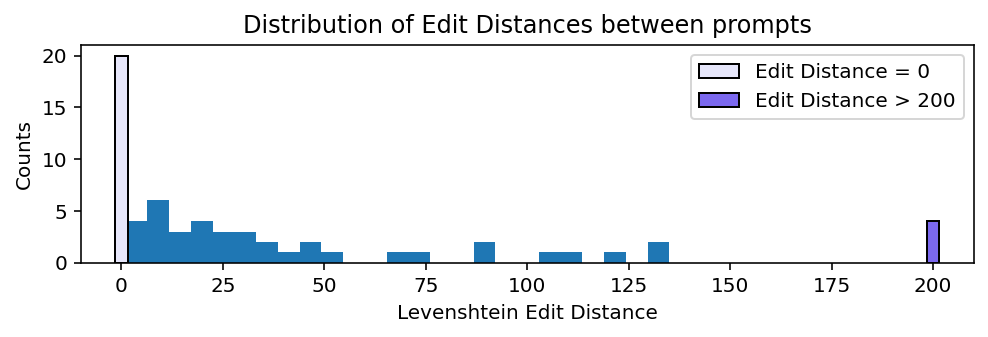}
\end{subfigure}
\vspace{-1em}
\caption{Distribution of the normalized edit distances between two consecutive prompts of the entire data, including prompts written for different tasks (top), and distributions of the normalized edit distances (middle) and the edit distances (bottom) between two consecutive prompts for the \textit{same task}. }
    \label{fig:tweak}
\vspace{-1em}
\end{figure}

After confirming that re-prompting behavior may signal an unsatisfactory \tc edit, we checked whether developers make efforts to receive a better \tc edit, by tweaking their prompts.
Similar to the identical re-prompting data collection, we considered a request as a re-prompting, if the request was within 10 minutes after the previous request, and the selected code for the requests was similar.
Then, to measure the differences between the two prompts, we computed the Levenshtein distance, the minimum number of characters to edit to change one string into the other, and the normalized Levenshtein distance by dividing it by the maximum length of the two prompts.

\Cref{fig:tweak} displays the result.
Out of $\sim$108000 pairs of prompts, we found $\sim$22000 (20\%) were identical re-prompts.
However, as it is possible that the two prompts are for different requests, such as [\textit{``rename this''}, \textit{``add null check''}], aside from the identical re-prompts, the edit distances cannot be directly interpreted.
Thus, we manually investigated $\sim$130 pairs of prompts that were requested from the same code region, and found that around 50\% of them were written for different tasks.
\Cref{fig:tweak}-middle and -bottom display the distributions of edit distances for the same task after excluding the 50\% of pairs written for different tasks.
The skewed distributions of edit distance show that developers rarely update their prompts when re-prompting, and tend to make small edits.
This is an impediment to using \tc most effectively, because the quality of prompts significantly influences the quality of the LLM response.

\subsection{Discussion}

Overall, we observed that developers often do not invest sufficient time in crafting effective prompts when using \tc.
Although documentation and the \tc user interface instructed users to provide specific details and necessary context, we found that few users did so when tweaking the prompt.
Developers typically resorted to manual editing or invested minimal effort in refining the prompt for a subsequent attempt when an initial, low-effort attempt with \tc yielded unsatisfactory code edits. 
This behavior suggests that there may be an `economics of prompting,' where developers weigh the costs of prompting against the expected benefits.
Thus, to improve the user experience and maximize the benefits of \tc, enhancing the underlying model performance alone is not sufficient; efforts must also focus on reducing users' cognitive load during prompting and on helping them write prompts more effectively~\cite{errica2024did, nam2024understanding, liu2023what}.

\section{Unsatisfactory Code Editing Dataset}
\label{sec:data}

To inform the mitigation strategies that help developers write more effective prompts for \tc, we aimed to understand current prompting practices and identify areas for improvement. 
We collected a dataset of real-world \tc usage, specifically focusing on developer requests that yielded \textit{not} satisfactory outcomes.

\subsection{User Submitted Unsatisfactory Usage}
\label{subsec:manualdata}
\tc has a feedback mechanism, where developers unhappy with the code edit suggestion from \tc can submit their desired code edit instead. This improved suggestion is bundled with the original prompt, code context, and original \tc suggested code edit upon submission. 
For these examples, the desired code edits most probably satisfy the user intent and expectations, as developers themselves need to deliberately create and submit the improved rewrite.
However, not all developers are motivated to invest time in submitting such rewrites; therefore, the number of these examples is quite limited.
After excluding the examples used for the fine-tuning of the \tc model, we had 52 examples left for this analysis.

\subsection{Automatically Extracted Unsatisfactory usage}
\label{subsec:autodata}
To collect more examples for the analysis, we used the \tc telemetry logs (see \Cref{sec:empirical}) to generate a dataset containing user requests that resulted in unsatisfactory code edits, along with the desired code edit.
As found in the previous section, we used identical re-prompting as an indicator of user dissatisfaction with the initial code edits suggested by \tc.

To obtain the desired code edit, we extracted a future snapshot of the code, assuming that regardless of how (un)satisfied the developers were, the task would have been resolved at some later point in time.
We consider the code state after 30 minutes, to leave enough time for developers to make their desired edits, especially for cases where the task requires finding additional information.
However, given that the edit time can vary significantly, depending on the task complexity and the user expertise, it was possible that code edits not relevant to the original \tc request could have been included. Developers could also potentially give up on finishing the task, or work on different tasks in the meantime.
This made manual verification necessary.

After collecting candidate \tc generated code edits, paired with the desired code edits, we applied filtering heuristics to make the manual verification process tractable, by excluding examples in which:
\begin{itemize}
    \item the original code (before the request) and the desired code edit are identical (\ie the user decided not to make the code edit),
    \item the \tc generated edit and the desired edit are identical (\ie the user was satisfied with the \tc suggestion),
    \item the difference between the original code and the desired code edit, or the \tc generated code edit is too long (the length of the unified diff is longer than 10,000 characters)
    \item there is no future snapshot (+30 minutes) available, because the user deleted the file, or
    \item the desired code edit only removes code from the original code. 
\end{itemize}
 This produced 496 examples, which were reviewed manually by two of the authors. 
We further filtered away examples---manually---where the edit included changes unrelated to the intent of the original developer prompt. This resulted in 107 high-quality examples after filtering.

\subsection{Full Dataset for Analysis and Solution Building}
Combining the data collected from the two sources resulted in 159 examples.
Among these, we used 111 examples (70\%) of the dataset for understanding developer prompting practices and issues, and we set aside 48 (30\%) to be used for testing the solution we would build based on our gained understanding.
Among these, we excluded examples that request code documentation only or simple renaming, where it is challenging to understand why the original \tc code edits were not satisfactory, and there is a subjective component involved (see Table E-Example 3 in \cite{artifact}).
This resulted in a training dataset of 72 examples, and a test dataset of 33 examples.
As the data were collected from real-world logs containing proprietary code, we cannot share the entire dataset, but we can only report the results and share examples.

\section{Phase 2: Understanding Common Issues of Developer Prompts}
\label{sec:erroranalysis}

\begin{figure}[tbp!]
    \centering
    \includegraphics[width=0.95\linewidth]{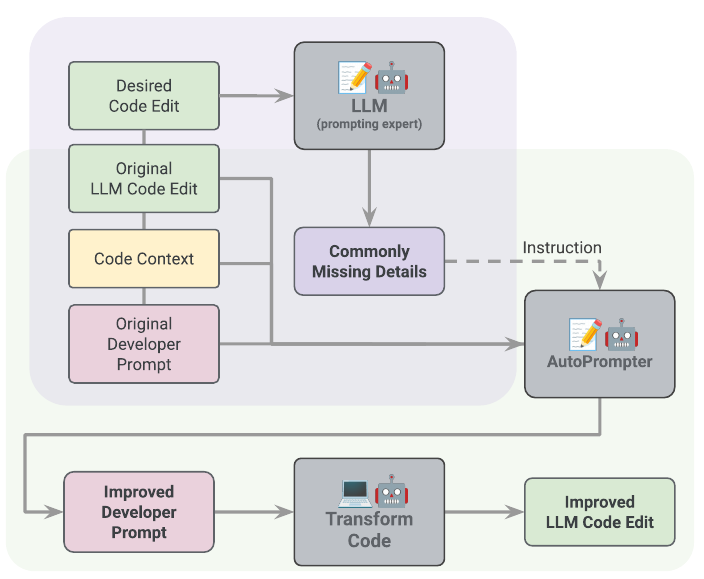}
    \caption{Overview of the developer prompt error analysis (purple) and automatic prompt enhancement (green). For error analysis, \gemini was given (1) the original user prompt, (2) the original \tc edit, (3) the code context, and (4) the desired code edit, to identify missing context from the original user prompt. For automatic user prompt enhancement, \tool was instructed with the learned missing context from the error analysis, and given (1) the original user prompt, (2) the original \tc edit, and (3) the code context. Then, \tc generates the code edit using the improved prompt.}
    \label{fig:overview}
\end{figure}

With the collected dataset, we conducted an error analysis of the developer requests that resulted in unsatisfactory code edits from \tc, to understand what would be required in a \textit{better prompt}.   
We leveraged an LLM\footnote{in our case, Gemini, but we could have used any of the popular LLMs.} to help us identify for each example the missing ingredients in a prompt, given prior knowledge of the complete code context, the prompt, and the desired edit.  
This is based on the premise that with such a \textit{better prompt}, \tc would likely have generated the desired solution.

\subsection{Error Analysis using an LLM}
\label{sec:attribution}

\begin{table*}[htbp!]
  \centering
  \caption{Summary of common issues with user prompts on 72 of our train examples.}
  \label{tab:codebook}
  \begin{tabular}{p{0.2\linewidth} p{0.08\linewidth} p{0.65\linewidth}}
    \toprule
    Code & \# &	Description\\
    \midrule
    
Specifics & 28 & Missing or under-specified specifications, for example, the prompt lacks value information \\
Operationalization plan & 28 & Missing or under-specified information on what operations need to be performed or what functions need to be used, or what edge case handling is needed to get to the desired code edit \\
Localization/scope & 6 & Missing or under-specified localization or scope of the change \\
Codebase context & 11 & Prompt itself did not provide the necessary background information for the model to be able to generate the desired code edits, including missing information about the broader system, existing codebase. Developers may have assumed that the model had all the necessary context. \\
User intent & 12 & Missing or under-specified user intent in the prompt; it is not clear what the developer's goal is or what type of code edit they want to achieve \\
    \bottomrule
  \end{tabular}
\end{table*}

To collect the common issues with user prompts (\ie what is missing from the prompt so that \tc gives the desired result),
we presented \gemini with: (1) the original user prompt, (2) the code context, (3) the original \tc edit, and (4) the desired code edit. Then we asked \gemini to produce (a) an explanation of what was missing from the original user prompt and (b) an improved prompt. 
See \Cref{appendix:qualprompt}~\cite{artifact} for the full system prompt used.

\subsubsection{Feasibility Testing}
\label{subsec:errorfeasibility}
As a sanity check to test whether an LLM can reason about the issues in the user prompts, we tested whether \tc can replicate the desired edits from \gemini-enhanced prompts, when provided with the desired edits to enhance the prompt.
Out of 72 examples, we found that 43 enhanced prompts (59.72\%) enabled \tc to replicate the desired edits. 
These numbers indicate that \gemini is able to pinpoint the issues with the user prompts well, and with all the provided information in the prompt, \tc is able to generate satisfactory outcomes for a fairly large number of cases.
At the same time, we found that not every unsatisfactory \tc edit is due to an incomplete user prompt. 
For example, cases that cannot be easily abstracted into a concise natural language prompt (\eg adding tens of import files without clear patterns across them) could not be improved by rewriting prompts even after seeing the desired edit.
We also discovered that in some examples, \tc provided better code edits than the desired code edits.
Here the developer may have accepted the \tc suggestion, had they seen it, but as we do not know the developer's original intent, they were not considered semantically identical.
Finally, the underlying model has limitations; it may not be able to generate the desired code edit even when presented with the perfect prompt.

\subsection{Qualitative Analysis}
We qualitatively analyzed the issues of user prompts for \tc extracted from the 72 examples in our dataset, using \textit{thematic analysis}~\cite{braun2006using}.
Our qualitative analysis consisted of several phases.
Three authors analyzed the original user prompts and the feedback provided by \gemini, which indicated what information was missing from the original prompt, using inductive coding~\cite{fereday2006demonstrating}. 
As we reviewed the data, we generated an initial set of categories to describe the different types of missing context and specifications. 
If we could not clearly identify which information was missing from the original prompt, we referred to the code context, the original \tc edit, and the desired code edit, which also helped mitigate any errors coming from potential hallucinations. 
We also wrote analytic memos during the discussion, including observations of patterns and potential connections among the categories. 

After analyzing 18 examples (25\% of the data), we sorted and consolidated the categories into a codebook (\Cref{tab:codebook}) with detailed definitions and examples.
We then revisited the 18 examples to ensure consistent code application. 
Then, two researchers separately categorized the remaining 54 examples using the codebook.
The first round of separate coding resulted in a Krippendorff's alpha score of 0.81, which shows good agreement.
Then, the ambiguous or difficult examples the two researchers did not agree on were discussed among the three researchers until agreement was reached.

\subsection{Results}

The analysis resulted in 5 codes of missing or under-specified information in user prompts for \tc.
\Cref{tab:codebook} shows the codebook~\cite{ando2014achieving}, including the code, the frequency of the codes, and the description.
Each example could be labeled with up to 2 codes.

\mysec{Specifics}
This code includes examples that lack specifics for the code edit, such as the variable name to use, the value to assign, the API call to use, or the type to use. 
For example, \tc produces a code edit that is different from the desired code edit given a user request with the prompt \textit{``fill result vector''}, because \textit{``The prompt did not specify the desired type of the `result` vector. It only mentioned filling a ``result vector,'' which is ambiguous.'' }
By using a prompt with a more complete specification, \textit{``Modify the [function] function to create a [type] named `result'.''}, \tc was able to generate the desired code edit.

\mysec{Operationalization Plan}
When there are multiple ways of implementing the same goal, if the user does not specify a particular solution, \tc may choose the most likely or probable solution. 
For example, given the user prompt \textit{``replace with read -r -p''}, \gemini pointed out that \textit{``The prompt lacked explicit instruction on the desired syntax for assigning the user input to the [variable1] variable. The prompt only specified replacing a hardcoded value with user input.'' }
With the improved prompt, \textit{``Modify the script to prompt the user for [variables], assigning the input directly to the respective variables using `read'.''}, \tc was able to replicate the desired code edit.
Including the operationalization plan can be critical when a user has a specific code modification in mind. 
However, in cases when developers want to make open-ended requests, such as \textit{``write javadoc''}, it might be less important \textit{how} \tc implements this change.

\mysec{Localization/Scope}
This code includes cases where the user prompt does not provide sufficient information about \textit{where}, including the location in the code context (e.g., lines) or scope (e.g., all or just one), the code edits should be made. 
Given the user prompt \textit{``surround values with \#{}''}, \gemini commented that \textit{``The prompt should explicitly state that all function calls within the selected lines need to be wrapped,''} and when improved as \textit{``Wrap all function calls within the [class] block's CSS variables with `\#{ }`.''}, \tc could replicate the desired edit.

\mysec{Codebase Context}
Codebase context refers to cases where the user assumes that \tc would know all the background knowledge and details about the codebase they are working on, in cases where \tc actually had no access to such information. 
This could happen because end users might not have a good understanding of how the underlying model is trained, which data are used, and what additional context is included as part of the request, which may have led the developers to assume a level of contextual awareness that the model does not possess.
For example, the desired code edit that \tc was not able to produce given \textit{``add [library]''} was successfully replicated with an improved prompt, \textit{``Add ``[path to the library]'' as a dependency to the [build] rule``.}

\mysec{User Intent}
A frequent example of this category is a user asking \tc to perform a rather vague and open-ended task.
For example, for a request with the original user prompt \textit{``make it simpler''}, \gemini pointed out that \textit{``The prompt did not specify the intended behavior when [variable1] is present in [variable2]. The model incorrectly interpreted ``make it simpler'' as removing the redundant assignment instead of preserving the original conditional logic with the necessary `continue' statement.''} 
The improved prompt generated by \gemini, \textit{``Refactor the conditional logic for assigning [variable3] to maintain the original behavior, adding a `continue' statement when [variable1] is in [variable2].'',} successfully replicated the desired code edit. 
When prompts lack user intent, compared to other issues, it becomes extremely difficult for \tc to produce the desired code edit, as it almost requires reading the minds of developers. 
Thus, if the desired code edit matches the most likely code edit to be made in general, given the code context, \tc can produce a good code edit, but if not, it is very challenging to achieve the desired code edit.

\mysec{Satisfactory Prompts}
We identified five examples from the dataset where the original user prompts appeared satisfactory without requiring further improvement, at least from the researchers' perspective. 
Because the dataset is a collection of user requests that resulted in unsatisfactory code edits, rather than a collection of inadequate prompts, it is reasonable that some of these unsatisfactory code edits are not attributable to the user prompts. 

\subsection{Feasibility of Automatic Analysis}
\label{subsec:autoqualitaitive}
In this study, we conducted a qualitative analysis following the traditional thematic analysis process, as this was the first attempt to understand user prompt issues for LLMs in the context of code editing.
However, this process of extracting insights from usage data could potentially be automated with LLMs, similar to other attempts in conducting qualitative research with LLMs~\cite{bano2023exploring, meng2024exploring}, particularly to reduce researcher effort and cost.
To assess the feasibility of such automation, we instructed an LLM to carry out the open coding and codebook construction.
We found the LLM-generated codebook to be largely similar to the one created by researchers (see \Cref{appendix:llmcodebook}~\cite{artifact}).
This result suggests the potential for applying automated analysis with low cost to understand user challenges in other tool contexts (\eg automatic program repair~\cite{rondon2025evaluating}) or domains (\eg code migration~\cite{nikolov202how}).

\section{Phase 3: Automatic User Prompt Enhancement}
\label{sec:autoprompter}
In the previous section, we identified categories of common issues with user prompts.
In this section, we tested the feasibility of automatically improving user prompts by exploiting this finding, and tested whether some of the missing or underspecified user requirements could be inferred from the code context.

\begin{table*}[!ht]
\caption{Example of prompt enhancement and the improved \tc code edits. \colorbox{added}{Green} indicates the added lines,  \colorbox{deleted}{red} indicates the deleted lines, and \colorbox{yellow}{yellow} indicates the changed lines.}
\label{tab:motivatingexample}
\begin{small}
\begin{tabular}{p{0.48\linewidth} p{0.48\linewidth}}
\toprule
\multicolumn{2}{l}{\textbf{Desired code edit extracted from the future code snapshot}}        \\ \midrule
\begin{minipage}[t]{\linewidth}
\vspace{-1em}
\begin{lstlisting}[style=normal]
hosts = [(*\MDELETE{]}*)

for host in existing_hosts(*\MDELETE{:}*)
    (*\MDELETE{name = \_HOST\_RE.search(host).group(1)}*)
    (*\MCHANGE{ip =}*) _HOSTNAME_RE.search(host).group(1)
    (*\MDELETE{if name not in new\_hosts and ip not in new\_ips:}*)
        (*\MDELETE{hosts.append(host)}*)
\end{lstlisting}
\end{minipage}
&
\noindent
\begin{minipage}[t]{\linewidth}
\vspace{-1em}
\begin{lstlisting}[style=normal]
hosts = [
    (*\MADD{host}*)
    for host in existing_hosts
    (*\MADD{if \_HOST\_RE.search(host).group(1) not in new\_hosts}*)
    (*\MCHANGE{and}*) _HOSTNAME_RE.search(host).group(1) (*\MADD{not in new\_ips}*) 
    (*\MADD{]}*)
\end{lstlisting}
\end{minipage}
\\ \midrule
\multicolumn{2}{l}{\textbf{Original developer prompt}: \textit{``simplify this''}}        \\ \midrule
\noindent
\begin{minipage}[t]{\linewidth}
\vspace{-1em}
\begin{lstlisting}[style=normal]
hosts = []
for host in existing_hosts:
\end{lstlisting}
\begin{lstlisting}[style=normal]
    (*\MDELETE{name = \_HOST\_RE.search(host).group(1)}*)
    (*\MCHANGE{ip =}*) _HOSTNAME_RE.search(host).group(1)
    (*\MDELETE{if name not in new\_hosts and ip not in new\_ips:}*)
        hosts.append(host)
			
hosts.append(_SECTION_RE.search(compute_section).group(2))
\end{lstlisting}
\end{minipage}
&
\noindent
\begin{minipage}[t]{\linewidth}
\vspace{-1em}
\begin{lstlisting}[style=normal]
hosts = []
for host in existing_hosts:
\end{lstlisting}
\begin{lstlisting}[style=normal]
    (*\MADD{if (host not in new\_hosts and}*)
        _HOSTNAME_RE.search(host).group(1)(*\MADD{ not in new\_ips):}*)

    hosts.append(host)
    
hosts.append(_SECTION_RE.search(compute_section).group(2)) 
\end{lstlisting}
\end{minipage}
\\ \midrule
\multicolumn{2}{l}{\textbf{\tool-enhanced prompt}: \textit{``Replace the loop with a more concise list comprehension for better readability.''}}        \\ \midrule
\begin{minipage}[t]{\linewidth}
\vspace{-1em}
\begin{lstlisting}[style=normal]
(*\MDELETE{hosts = []}*)
(*\MDELETE{for host in existing\_hosts:}*)
    (*\MDELETE{name = \_HOST\_RE.search(host).group(1)}*)
    (*\MDELETE{ip = \_HOSTNAME\_RE.search(host).group(1)}*)
    (*\MDELETE{if name not in new\_hosts and ip not in new\_ips:}*)
        (*\MDELETE{hosts.append(host)}*)
\end{lstlisting}
\end{minipage}
&
\noindent
\begin{minipage}[t]{\linewidth}
\vspace{-1em}
\begin{lstlisting}[style=normal]
(*\MADD{hosts = [host for host in existing\_hosts}*)
        (*\MADD{if \_HOST\_RE.search(host).group(1) not in new\_hosts}*)
        (*\MADD{and \_HOSTNAME\_RE.search(host).group(1) not in new\_ips]}*)
\end{lstlisting}
\end{minipage}
\\ \midrule

\end{tabular}
\end{small}
\end{table*}

\subsection{Approach to Prompt Enhancement}
\label{sec:enhancement}
To test the feasibility of, and the extent to which, automatically enhancing developer prompts, we built and evaluated \tool using an LLM (see \Cref{fig:overview}).
Given (1) the original developer prompt, (2) \tc suggested edit generated with the original developer prompt, and (3) the code context provided by the developer, \tool was instructed to check whether the given developer prompt misses or under-specifies any of the key information, and to infer them from the given code context to enhance the prompt.
We also instructed \tool to declare if it cannot confidently infer an improved prompt because of a lack of information, and to explicitly request additional context instead of giving the improved prompt.

As \gemini is inferring the missing specifications from the code context, it is very likely that there can be multiple probable ways of further specifying the developer prompt.
For example, when a developer asks to \textit{``document''}, it can be \textit{``improve documentation to be more concise''} (user intent) but also \textit{``write docstrings for every function''} (localization/scope).
To make sure that \tool generates diverse prompts that can cover multiple possible specification options, we instructed it to provide missing/unspecified instructions given the categories we found in the previous section as a reference, and to improve prompts considering each of those categories. The prompt used can be found in \Cref{sec:prompt}~\cite{artifact}.

\subsection{Evaluation}
\label{subsec:eval}

\subsubsection{Dataset}
We used 33 user requests to \tc that resulted in unsatisfactory code edits to evaluate the quality of enhanced code edits.
See \Cref{sec:data} for further details.
We note that the goal of this evaluation is to show the feasibility of automatically improving the developer prompt for programming and to demonstrate how such a tool can be evaluated, rather than claiming that \tool is the ideal approach for developer prompt enhancement.

\subsubsection{Metrics}
To evaluate the quality of code edits, we used multiple metrics:

\mysec{ChrF (Character-level F-score)~\cite{popovic2015chrf}}
Computes an $F_1$ score based on the proportion of character-level $n$-grams.

\mysec{Gestalt pattern matching} 
Also known as Ratcliff/Obershelp matching. Recursively computes the lengths of the longest common substrings of two strings and normalizes the length of overlapping regions by the total length of both strings. 

\mysec{Autorater~\cite{zheng2023judging}}
A complement to text-based similarity scores, which can suffer from syntactic differences, measures the semantic similarity between code edits using an LLM-as-a-judge. 
Given the original developer prompt, the \tool enhanced prompt, and the desired code edits, \gemini evaluates the similarity; \ie how similar the code edits are to the desired code edit.

\mysec{Human evaluation} 
Two authors separately evaluated the improved code edits by comparing them to the ground-truths and the original user prompts.
Each example was labeled as 1 when the improved code edit was semantically equivalent to the ground truth, considering the original user prompt and the code context, and 0 if not.
After a separate evaluation, we found a substantial agreement by achieving a 0.68 Cohen's Kappa score.
For cases where the two authors did not agree, we averaged the two scores.

We note that none of the metrics can accurately represent the actual quality of code edit performance, but they are complementary to each other, and the combination should be interpreted altogether.

The other popular text-based metrics, such as BLEU~\cite{papineni2002bleu}, are not suitable for assessing the quality of code edits.
We also considered measuring the CodeBLEU~\cite{ren2020codebleu} scores for the evaluation.
However, as CodeBLEU is language-specific in implementation and given the variety of languages our benchmark contains, we decided that extending the CodeBLEU implementation was not practical.

\subsubsection{Limit Study}
\label{subsec:limit}

As discussed in \Cref{subsec:errorfeasibility}, not every code edit benefits from prompt enhancement.
Recognizing that improvement is not expected for all test set examples, we established a theoretical performance \textit{limit} as a reference in \Cref{subsec:toolresults}.
To determine this limit, similar to the feasibility test (\Cref{subsec:errorfeasibility}), we provided \tool not only with the regular input but also with the desired final code edit, which ensured \tool received the maximum possible information.
Although evaluating against the full test set might seem unfair because some examples fall outside the scope of our prompt enhancement approach, we decided to keep them as we believe the current setting represents real-world usage.

\subsection{Results}
\label{subsec:toolresults}

\begin{table}[tbp!]
  \centering
  \caption{Evaluation results. Original prompt reports the \tc edits generated from the original user prompt. Self-selection, farthest, and llm report the results of different selection strategies. Limit reports the maximum boundary (Section 6.2.3).  }
  \label{tab:eval}
  \begin{tabular}{p{0.23\linewidth} p{0.12\linewidth} p{0.12\linewidth} p{0.13\linewidth} p{0.13\linewidth}}
    \toprule
                        & ChrF              & Gestalt           & Autorater         & Human.  \\ \midrule
    original            & 0.652             & 0.619             & 0.545             & 0 \\ \midrule 
    none                & 0.646             & 0.609             & 0.500             & 0.167 \\ \midrule 
    single               & 0.664             & 0.628             & 0.605            & 0.212  \\
    self-selection      & 0.621             & 0.585             & 0.531             & 0.167 \\
    farthest            & 0.656             & 0.634             & \textbf{0.673}    & \textbf{0.272} \\
    llm                 & \textbf{0.699}    & \textbf{0.661}    & 0.638             & 0.259 \\ \midrule
    Limit               & 0.771             & 0.720             & 0.719             & 0.367\\
    \bottomrule
  \end{tabular}
\end{table}

\Cref{tab:eval} summarizes the results.
In interpreting the data, we focused on human evaluation results, as we found that text-based similarity scores often do not correlate with human evaluation, especially when the differences are small.
Moreover, text-based metrics often give unsatisfactory results in the case of syntactic differences that do not matter semantically.
The semantic human evaluation count is 0 by design, as the dataset was built to only include unsatisfactory \tc code edits, \ie semantically different from the desired code edit.

As two baselines, we used \texttt{original} and \texttt{none}. 
\texttt{original} displays the scores of original \tc edits, based solely on the original user prompts. 
\texttt{none} is based on the scores of \tc response generated only with \gemini prompt enhancement, but without including \Cref{tab:codebook} as part of the instruction. The difference between this and other scores thus indicates the usefulness of instructing \tool to consider the common issues of user prompting.
We have multiple versions of \tool, where the simplest is \texttt{single}, which asks \gemini to produce one enhanced prompt given the information.
Also, recall from Section~\ref{sec:enhancement} that we ask \gemini to produce multiple prompt variants.
Given multiple prompts, each one improved with a different strategy, we employed various methods to select the best improved prompt to use.
In \Cref{tab:eval}, \texttt{self-selection}, \texttt{farthest}, and \texttt{llm} show the performance of different selection approaches.
First, for (\texttt{self-selection}), we instructed the \tool directly to select a prompt that is most likely to elicit the desired code edit.
We also tried choosing a prompt for which \tc produces the most distinctive code edit from the original code edit (\texttt{farthest}).
This is based on the idea that if the developer is not satisfied with the original code edit, the desired code edit is more likely to be found within code edit candidates that differ from the unsatisfactory code edit than within those similar to it.
Finally, we used an LLM to select the best improved \tc code edit, given the original developer prompt (\texttt{llm}).

Overall, the evaluation shows the feasibility of automatically improving the developer prompt:
9 out of 33 (27\%) examples were improved by \tool, which indicates that it is often feasible to infer missing specifications from the given code context.  
This seems like a good result, especially when considering 37\% as the limit (as computed by the process in ~\Cref{subsec:limit}) for this test set.
Among the \tool variations, \texttt{self-selection} appears to perform the worst; this could indicate that it might not be easy for an LLM to tell whether the given prompt meets the user intent or not, without reviewing the code edits generated by it.
Recall also from \Cref{subsec:limit} that some of the examples may not be amenable to this process of prompt enhancement; and, finally, \tc also has limits in terms of the power of the model itself in interpreting the prompt and carrying out a desired edit.

\section{Threats to Validity}

\subsubsection*{Telemetry logs analysis for \tc}
The use of one month of data may be insufficient for broad generalization, particularly given the rapid evolution of AI tools at \company, and the corresponding adaptation of developers. 
Furthermore, updates to the LLMs supporting the AI tools, which occur during tool evolution, might influence developer satisfaction and behavior. 
However, we had to tolerate these limitations to balance the complexity of data collection, privacy considerations, and the scale of analysis. 
Given the substantial size of \company's internal developer base and the widespread adoption of \tc among developers over an extended period, we believe that our dataset is representative, but should be interpreted considering the use case, user population, and the scope of features.

\subsubsection*{Qualitative Analysis for Common Issues in Developer Prompts}
The error analysis is subject to the typical threats to validity encountered in other qualitative analyses.
While analyzing data qualitatively allows us to engage in-depth and distill knowledge from previously unexplored areas, we cannot guarantee that the findings generalize beyond the sampled examples without further validation. 
Specifically, because complex and lengthy examples were excluded to facilitate practical manual review, our sample may not encompass all possible prompts used for code editing with an AI tool. 
However, given that \company covers diverse application domains and developers with varying expertise in programming, domain knowledge, and AI tools, we believe the sampled examples reasonably represent a diverse range of use cases for code editing tasks.

\subsubsection*{Correlating in-IDE and \tc events} We already discussed at length in \Cref{sec:empirical} the issues in achieving perfect alignment between IDE logs and \tc logs, and this could hinder some precision in our analysis; however, with the scale, we believe this concern is mitigated. 

\subsubsection*{Specific to \tc} Our experiments with improving prompt effectiveness have all been in the context of \tc.  We believe that the ideas will carry over to similar tools where prompts are used to make code edits (\eg Cursor~\cite{cursor}).

\section{Discussion \& Conclusion}

In this paper, we have studied how we can improve developer prompts so that \tc produces code edits that more closely align with developer expectations and intentions.

Our approach is to analyze the common reasons that the prompts themselves could be lacking.  We did this by leveraging an LLM and asking it to articulate the prompt enhancements that would be needed to get \tc to produce edits matching the given desired edit.  We collated this information into a codebook that captures five common classes of missing information: missing some specific information, missing an operationalization plan, faulty localization/scope, missing codebase context, or missing intent.

This analysis then enabled us to create AutoPrompter, a tool that proactively adds details to the prompt, without knowing the ground truth, to get \tc to produce a more satisfactory edit from a given developer prompt.  
This idea seems to have shown promise, as we show in our test set: 27\% of the cases that were not achieving a satisfactory result with \tc now accomplished a satisfactory result with the automatically enhanced prompt.

Although we focused on testing the feasibility of automatically enhancing user prompts in this paper, there is a lot of room for future work.
First, the identified common issues with user prompts can be used to guide the training strategies of the underlying LLM for \tc, such that \tc can infer such missing information even with the original, under-specified user prompts.
Using an automated analysis of usage data with LLMs (see \Cref{subsec:autoqualitaitive}), this process of learning from user struggles can be generalized to other LLM-powered tools in identifying gaps.

Another promising usage of \tool would be to use it for the automatic curation of training data from the usage logs.
\tc usage data can be very useful for training, as it offers negative examples (original \tc edit) alongside ground-truths (user ``rewrites'') given user prompts. 
However, only a small, human-verified subset of user-submitted ground truths (see \Cref{subsec:manualdata}) has been used, due to the challenges in extracting quality ground-truths, discussed in \Cref{subsec:autodata}.
The core idea is to include the \tool-improved \tc edit (edit generated using the \tool-enhanced user prompt) as ground-truths, \textit{only if} they are identical to a future snapshot. 
Because \tool only augments the prompt based on the provided context, the improved \tc edit captures the original user intent without adding requirements not mentioned in the original prompt. 
The identity of a future code snapshot also shows implicit user confirmation that the improved edit was desired. 

Finally, one potential benefit of \tool, not discussed in this paper as it is out of scope, is the increased transparency and control it offers users.
As an enhanced prompt is generated to obtain an enhanced code edit from \tc, when presented, users can easily verify whether their intent is accurately described in the prompt. 
This prompt can then be easily updated by the user if it contains errors, affording them direct control over the input to \tc.
In addition, presenting these improved prompts to developers could help them learn to write more effective prompts themselves~\cite{ma2024engineer}, possibly enabling them to understand the benefits of adding more context. This understanding, in turn, could lead to better code edits from \tc with less accumulated effort.

\newpage
\bibliographystyle{ACM-Reference-Format}
\bibliography{reference}

\balance

\appendix
\onecolumn

\section{System Instructions for Error Analysis}
\label{appendix:qualprompt}

\begin{prompt}
You are an expert Prompt Engineer specializing in code editing prompts for a Code LLM. Your task is to help software engineers improve their prompts to achieve desired code modifications.

You will receive the following inputs:

1.  \textbf{Current Prompt:} The engineer's initial prompt to the Code LLM.

2.  \textbf{Code Context:} The complete code snippet being edited.

3.  \textbf{Selected Lines:} The specific lines of code the engineer highlighted in their request.

4.  \textbf{Current Model Diff:} The difference (in unified diff format) between the original code and the current, \textit{undesired} output from the Code LLM.

5.  \textbf{Golden Response Diff:} The difference (in unified diff format) between the original code and the \textit{desired} (golden) output.

Your task is to analyze these inputs and generate a report in Markdown format, containing the following sections:

1.  \textbf{Analyze Differences:} Carefully examine the *Current Model Diff* and the *Golden Response Diff* to understand the discrepancy between the current and desired outputs.

2.  \textbf{Identify Missing Specification:} Determine what crucial information is *missing* from the original prompt that might be preventing the Code LLM from generating the desired output.

3.  \textbf{Create Improved Prompt:}  Construct a new prompt that is clear, specific, concise, and error-free.

4.  \textbf{Request More Information (If Necessary):} If you cannot confidently infer the user's intention or create an improved prompt due to insufficient information, provide a clear request for additional context instead of the improved prompt. This request should also be concise.
\end{prompt}

\section{Example Output Produced for Error Analysis}

\begin{prompt}
\textbf{Diff Analysis}

The current model diff replaces the nested loops iterating through [variable1] and calculating min/max values with a more concise approach using [function1] and [function2].  However, the golden response diff maintains the nested loop structure but uses [function3] and [function4] within the loop for a more efficient calculation of min/max values.  The current model's approach is potentially less efficient for large datasets. The golden response is more efficient and avoids unnecessary array creation.
\newline
\textbf{Missing Specification}
\\
-  Desired method for min/max calculation: The prompt ``simplify'' is too vague.  It doesn't specify whether to maintain the original nested loop structure while improving efficiency or to use a completely different, potentially less efficient, approach. This should be provided by the user.

\textbf{Improved Prompt}

Refactor [function5] to efficiently compute min/max values using [function3] and [function4] within the existing nested loop structure.

\textbf{Request for More Information (If needed)}

Please clarify whether you prefer a solution that maintains the original nested loop structure (as in the golden response) or allows for a different approach (as in the current model response).  If the former, confirm the intent is to improve efficiency, not change the algorithm.

\end{prompt}

\newpage

\section{LLM-generated Codebook}
\label{appendix:llmcodebook}

\begin{table*}[!ht]
\begin{tabular}{p{0.2\linewidth}  p{0.75\linewidth}}
\toprule
Code & Description \\ \midrule
Missing Context & The prompt lacks necessary background information or surrounding code context for Code LLM to understand the intended functionality or modification. This includes missing information about the broader system, existing code, or overall goals. \\
Missing Action & The prompt does not clearly specify what action needs to be taken (e.g., add, modify, delete, refactor) or the specific operation to be performed (e.g., sort, filter, convert). This includes missing verbs or insufficiently specific verbs. \\
Missing Object/Target & The prompt does not clearly identify the target of the action. This includes missing or ambiguous references to variables, functions, classes, components, files, data structures, or specific parts of a larger structure. \\
Missing Value/Specification & The prompt lacks detail about a required value, such as constant values, data types, strings, regular expressions, paths, URLs, configuration settings, build details, parameter or return types, or API-specific information. \\
Missing Logic/Condition & The prompt does not specify the conditional logic or control flow under which code should execute or behave. This includes missing if conditions, switch cases, loop conditions, error handling logic, and algorithm specifications. \\
Missing Style/Format & The prompt does not specify desired coding style, formatting preferences, or response presentation. \\
Missing Import/Dependency & The prompt fails to specify necessary imports, includes, or build dependencies required for the code to function correctly. \\
\bottomrule
\end{tabular}
\end{table*}

\section{System Instruction for Automatic Prompt Enhancement}
\label{sec:prompt}
\begin{prompt}
You are an expert Prompt Engineer specializing in code editing prompts for a Code LLM. Your task is to help software engineers improve their prompts to achieve desired code modifications. You will receive the following inputs:

1.  \textbf{Current Prompt:} The engineer's initial prompt to the Code LLM.

2.  \textbf{Code Context:} The complete code snippet being edited.

3.  \textbf{Selected Lines:} The specific lines of code the engineer highlighted in their request.

4.  \textbf{Current Model Diff:} The difference (in unified diff format) between the original code and the current, *undesired* output from the Code LLM.

Your task is to analyze these inputs and generate feedback in Markdown format, containing the following sections:

1.  \textbf{Identify Missing Specification:} Use the following rubric, analyze the user prompt against *each* code in the codebook. Ask yourself, "Does this prompt exhibit this type of missing information, that would prohibit the Code LLM from producing the desired output?"

[\Cref{tab:codebook}]

2.  \textbf{Generate Improved Prompts:} Suggest revised versions of the user prompt (each under 50 words) that address each of the identified weaknesses above. For each improved prompt, if inferrable, include the missing information to help the Code LLM make the desired change. Only suggest improved prompts for selected missing specifications, and do not suggest improved prompt if none were selected.

3.  \textbf{Recommend the Best Prompt:}  Evaluate the revised prompts and select the one you believe is most likely to elicit the desired code edit from the Code LLM.  Explain your reasoning.

4.  \textbf{Request More Information (If Necessary):} If you cannot confidently infer the user's intention or create an improved prompt due to insufficient information, provide a clear request for additional context instead of the improved prompt.  This request should be concise and specifically address the missing information needed.
\end{prompt}

\newpage
\section{Unsatisfactory \tc Code Edit Examples}

\begin{table*}[!ht]
\caption{Examples of unsatisfactory \tc code edits. Original Code Edits are those generated with the original user prompt (written above each block), and the Desired Code Edits are the code edits users expected \tc to generate with the original user prompt. \colorbox{added}{Green} indicates the added lines, \colorbox{deleted}{red} indicates the deleted lines, and \colorbox{yellow}{yellow} indicates the changed lines.}
\label{tab:examples}
\begin{tabular}{p{0.05\linewidth}  p{0.45\linewidth} p{0.45\linewidth}}
\toprule

\multicolumn{3}{c}{\textbf{Example 1}: ``Use ?:''}        \\ \midrule
\begin{minipage}[t]{\linewidth}
\footnotesize
Original
[705:709]
\end{minipage} &
\noindent
\begin{minipage}[t]{\linewidth}
\vspace{-1em}
\begin{lstlisting}[style=normal]
// Merge and split can happen consecutively on the \ 
        same dimension, e.g.,
// f32[1024,256] to f32[128,2048] can be considered \
        that 1024 gets split into
// 128 and 8, but 8 then gets merged with 256. \ 
        We use stacks to make
// supporting such cases easy.
\end{lstlisting}
\end{minipage}
&
\noindent
\begin{minipage}[t]{\linewidth}
\vspace{-1em}
\begin{lstlisting}[style=normal]
// Merge and split can happen consecutively on the \ 
        same dimension, e.g.,(*\MADD{?:}*)
// f32[1024,256] to f32[128,2048] can be considered \ 
        that 1024 gets split into
// 128 and 8, but 8 then gets merged with 256. \ 
        We use stacks to make
// supporting such cases easy.
\end{lstlisting}
\end{minipage}
\\ \midrule
\begin{minipage}[t]{\linewidth}
\footnotesize
Desired
[838:849]
\end{minipage} &
\noindent
\begin{minipage}[t]{\linewidth}
\vspace{-1em}
\begin{lstlisting}[style=normal]
for (int 64_t i = sharding.TiledDataRank();
    i < sharding.tile_assignment().num_dimensions(); ++i) {
  (*\MDELETE{if (i == sharding.SubgroupReplicationDim()) \{}*)
    (*\MDELETE{target\_tile\_assignment\_dimensions.push\_back(1);}*)
  (*\MDELETE{\} else \{}*)
(*\MDELETE{  }*)  target_tile_assignment_dimensions.push_back(

        sharding.tile_assignment().dim(i));
      (*\MDELETE{\}}*)
  }
\end{lstlisting}
\end{minipage}
&
\noindent
\begin{minipage}[t]{\linewidth}
\vspace{-1em}
\begin{lstlisting}[style=normal]
for (int64_t i = sharding.TiledDataRank();
    i < sharding.tile_assignment().num_dimensions(); ++i) {
    
    
    
    target_tile_assignment_dimensions.push_back(
        (*\MADD{i == sharding.SubgroupReplicationDim()}*)
            (*\MADD{? 1}*)
            (*\MADD{:}*) sharding.tile_asignment().dim(i));
    }
\end{lstlisting}
\end{minipage}
\\ \midrule
\multicolumn{3}{c}{\textbf{Example 2}: ``make it succint''}        \\ \midrule
\begin{minipage}[t]{\linewidth}
\footnotesize
Original
[119:125]
\end{minipage} &
\noindent
\begin{minipage}[t]{\linewidth}
\vspace{-1em}
\begin{lstlisting}[style=normal]
elif self.batch_size:

    (*\MCHANGE{  }*)if self.batch_drop_remainder:
(*\MDELETE{    }*)        return len(self._root_map_ds)  // self.batch_size
(*\MDELETE{    }*)    else:
(*\MDELETE{    }*)        return len(self._root_map_ds) // self.batch_size + 1
\end{lstlisting}
\end{minipage}
&
\noindent
\begin{minipage}[t]{\linewidth}
\vspace{-1em}
\begin{lstlisting}[style=normal]
elif self.batch_size(*\MADD{ is None}*):
    (*\MADD{return len(self.\_root\_map\_df)}*)
(*\MCHANGE{el}*)if self.batch_drop_remainder:
    return len(self._root_map_ds)  // self.batch_size
else:
    return len(self._root_map_ds) // self.batch_size + 1
\end{lstlisting}
\end{minipage}
\\ \midrule
\begin{minipage}[t]{\linewidth}
\footnotesize
Desired
[119:125]
\end{minipage} &
\noindent
\begin{minipage}[t]{\linewidth}
\vspace{-1em}
\begin{lstlisting}[style=normal]
(*\MDELETE{el}*)if self.batch_size:
    (*\MDELETE{if self.batch\_drop\_remainder:}*)
        (*\MDELETE{return len(self.\_root\_map\_ds) // self.batch\_size}*)
    (*\MDELETE{else:}*)
        return len(self.\_root_map\_ds) // self.batch\_size + (*\MCHANGE{1}*)

\end{lstlisting}
\end{minipage}
&
\noindent
\begin{minipage}[t]{\linewidth}
\vspace{-1em}
\begin{lstlisting}[style=normal]
if self.batch_size:



    return len(self._root_map_ds)  // self.batch_size + (*\MCHANGE{(}*)
        (*\MADD{not self.batch\_drop\_remainder}*)
    (*\MADD{)}*)
(*\MADD{else:}*)
    (*\MADD{return len(self.\_root\_map\_ds)}*)
\end{lstlisting} 
\end{minipage}
\\ \midrule
\multicolumn{3}{c}{\textbf{Example 3}: ``Add docstring for frontCamera''}        \\ \midrule
\begin{minipage}[t]{\linewidth}
\footnotesize
Original
[59:60]
\end{minipage} &
\noindent
\begin{minipage}[t]{\linewidth}
\begin{lstlisting}[style=normal]





 public boolean frontCamera = false;
\end{lstlisting}
\end{minipage}
&
\noindent
\begin{minipage}[t]{\linewidth}
\vspace{-1em}
\begin{lstlisting}[style=a]
/**
* Flag indicates if we are running the APK for front \  
    camera calibration. If this is set to true,
* the calibration APK will adjusts its text for front\ 
    camera calibration mode.
*/
\end{lstlisting}
\begin{lstlisting}[style=normal]
 public boolean frontCamera = false;
\end{lstlisting}
\end{minipage}
\\ \midrule
\begin{minipage}[t]{\linewidth}
\footnotesize
Desired
[59:60]
\end{minipage} &
\noindent
\begin{minipage}[t]{\linewidth}
\vspace{-1em}
\begin{lstlisting}[style=normal]


 public boolean frontCamera = false;
\end{lstlisting}
\end{minipage}
&
\noindent
\begin{minipage}[t]{\linewidth}
\vspace{-1em}
\begin{lstlisting}[style=a]
/** Indicates if the front camera should be used \
    for calibration. */
\end{lstlisting}
\begin{lstlisting}[style=normal]
 public boolean frontCamera = false;
\end{lstlisting}
\end{minipage}
\\ \midrule
\end{tabular}
\end{table*}

\newpage

\section{Additional Prompt Improvement Examples}

\begin{table*}[!ht]
\caption{Example of user prompt enhancement. }
\begin{tabular}{p{0.1\linewidth} p{0.85\linewidth}}
\toprule
\multicolumn{2}{l}{\textbf{Example 1}}        \\ \midrule
Original & adjust numbers starting from 61 \\
Enhanced& Increment the numbers in the enum [variable1] starting from `[variable2] = 61' to maintain sequential order. \\  \midrule
\multicolumn{2}{l}{\textbf{Example 2}}        \\  \midrule
Original & can you make the default [variable1] the same as the default [variable2]? \\
Enhanced & Set the default value of `[variable1]' to be the same as the value of `[variable2]', which is `[variable3]'.\\ \midrule
\multicolumn{2}{l}{\textbf{Example 3}}        \\ \midrule
Original & pass this in as an argument \\ 
Enhanced & Pass `[variable1]' as an argument to the `[function]' function. \\ \midrule
\multicolumn{2}{l}{\textbf{Example 4}}        \\ \midrule
Original & Change sass variables into css variables \\ 
Enhanced & Convert the selected lines to use CSS variables, referencing the Sass variables from `[variable1]' using the `[function1]' function. For example, `[name2]: [function1]([name2])'. \\ \midrule
\multicolumn{2}{l}{\textbf{Example 5}}        \\ \midrule
Original & Add default initialzed values to all members this class \\
Enhanced & Add default initialized values to all members of this class. Use `0' for `int' and `double', and `nullptr' where appropriate. Initialize vectors as empty. \\ 
\bottomrule
\end{tabular}
\end{table*}

\end{document}